\documentstyle[preprint,aps]{revtex}

\newcommand{\vp}{{\bf p}}
\newcommand{\al}{\alpha}
\newcommand{\bt}{\beta}

\newcommand{\be}{\begin{equation}}
\newcommand{\ee}{\end{equation}}
\newcommand{\ba}{\begin{eqnarray}}
\newcommand{\ea}{\end{eqnarray}}

\begin{document}

\title{Generalized Boltzmann 
equation for a trapped Bose-condensed gas using 
the Kadanoff-Baym formalism \footnote{This paper was presented at the 
workshop on ``Kadanoff-Baym Equations'' held in Rostock, Sept. 20-24, 
1999. It will be published in ``Progress in Nonequilibrium 
Green's functions", M. Bonitz (Ed.), 
(World Scientific, Singapore, 2000).}}

\author{M. Imamovic-Tomasovic and A. Griffin}

\address{Dept. of Physics, University of Toronto, 
Toronto, Ontario, Canada M5S 1A7 } 

\newcommand{\r}{{\bf r}}
\newcommand{\up}{\stackrel{<}{>}}
\newcommand{\down}{\stackrel{>}{<}}
\newcommand{\co}{({\bf R},T)}
\newcommand{\cog}{(\r, {\bf R}, T)}
\newcommand{\cof}{({\bf p},{\bf R},T)}
\newcommand{\nee}{\tilde{\epsilon}_{p}}
\def\pomrt{(\vp,\omega;{\bf R},T)}
\maketitle
\abstract{Using the Kadanoff-Baym non-equilibrium 
Green's function formalism, we derive kinetic 
equations for the non-condensate atoms 
at finite temperatures which include the effect of 
binary collisions between atoms. The effect of 
collisions is included using the second-order 
self-energy given by the Beliaev (gapless) 
approximation. We limit our discussion to finite 
temperatures where we can use the single-particle 
Hartree-Fock spectrum for the excited atoms. 
In this limit, we can neglect the off-diagonal 
propagators ($\tilde{g}_{12}$ and $\tilde{g}_{21}$).
As expected, this leads to the kinetic equations 
and collision integrals used in recent work by 
Zaremba, Nikuni, and Griffin (ZNG) \cite{zngjltp}.  
We also derive a consistent equation of motion 
for the condensate wavefunction, involving a 
finite-temperature generalization of the 
well-known Gross-Pitaevskii equation which 
includes a dissipative term, as well 
as the mean field of the non-condensate. }

\section{Introduction}
In a trapped, weakly-interacting Bose gas at 
temperatures low compared to the Bose-Einstein 
transition temperature $T_{BEC}$, the fraction of 
atoms that are excited out of the condensate is 
very small (only a few percent in contrast to 
90\% in liquid He4). As a result, the dynamics 
of the trapped Bose gas is completely described 
by an equation of motion for the macroscopic 
wavefunction $\Phi({\r},t)$. This is the famous 
time-dependent {\it Gross-Pitaevskii equation}
(GP) \cite{strtheory}
\be
i\frac{\partial \Phi({\r},t)}{\partial t}
=\left[-\frac{1}{2m}\nabla_{\r}^{2}
+U_{ext}({\r})+gn_{c}({\r},t)\right]\Phi({\r},t),
\label{eq:gpeq}
\ee
where $n_{c}({\r},t)=|\Phi({\r},t)|^{2}$ is the 
non-equilibrium density of the atoms in the condensate 
and $U_{ext}({\r})$ is a parabolic trap potential. 
In this article, we set $\hbar=1$. For a discussion 
of the properties of a dilute Bose gas at low 
temperatures, the two-body interaction 
$v({\bf r}-{\bf r'})$ can be treated using 
the {\it s}-wave approximation: 
\be
v({\bf r})=g\delta({\bf r}), \hspace{5mm} g=4\pi a/m. 
\label{eq:continteraction}
\ee
One can see that the GP equation describes the 
motion of the condensate moving in the Hartree 
mean-field produced by the other atoms in the condensate.
The GP equation is a closed equation for $\Phi({\r},t)$. 
It provides a very accurate description of the static
and dynamic properties of a trapped Bose gas at low 
temperatures $T\leq 0.4 T_{BEC}$, as confirmed by many 
experiments in the last few years \cite{strtheory}.

At finite temperatures, however, the number of atoms 
thermally  excited out of the condensate becomes 
significant and the GP theory is no longer valid. 
The simplest way to include the effect of the excited 
atoms on the condensate is to add the additional 
Hartree-Fock mean field  $V_{HF}=2g\tilde{n}({\r},t)$ 
produced by the non-condensate atoms (here $\tilde{n}$ 
is the local non-condensate density). One 
immediately sees this new equation is no longer 
closed since it involves the dynamics of 
the non-condensate atoms. 

To find the time-dependent non-condensate density, 
we need a {\it quantum Boltzmann equation} for the 
distribution function of the non-condensate atoms 
$f({\bf p},\r,t)$. A simple kinetic equation for a 
trapped Bose gas which has been used in recent studies  
\cite{zngjltp,stoofjltp} can be written in the following 
Boltzmann-like form 
\be
\left[\frac{\partial}{\partial t}+\frac{\bf p}{m}
\cdot \nabla_{\r}-\nabla_{\r}U({\r},t)\cdot 
\nabla_{\bf p}\right]f({\bf p},\r,t)
=\left[\frac{\partial f({\bf p},\r,t)}
{\partial t}\right]_{coll}. \label{eq:kinht}
\ee 
Here, the thermally excited atoms are assumed to be well 
described by the single-particle spectrum 
$\frac{p^{2}}{2m}+U({\r},t)$, where
\be
U({\r},t)\equiv U_{ext}(\r)+2g\left[n_{c}({\r},t)
+\tilde{n}({\r},t)\right]
\label{eq:effectivefield}
\ee
includes the self-consistent Hartree-Fock dynamic mean field
involving the {\it total} time-dependent local density 
$n({\r},t)$. The right-hand side of (\ref {eq:kinht})  
describes the effect of collisions between atoms on 
the evolution of the distribution function 
$f({\bf p},\r,t)$. In trapped Bose-condensed gases, 
it is the sum of two parts
\be
\left[\frac{\partial f}{\partial t}\right]_{coll}
=C_{12}[f]+C_{22}[f].
\label{eq:collint}
\ee
Here, $C_{22}$ denotes the part of the collision 
integral that describes two-body collisions between 
non-condensate atoms. Above $T_{BEC}$, this is the 
only term present. $C_{12}$ describes collisions 
between non-condensate atoms which involve 
{\it one condensate atom}. The role of $C_{12}$ is 
crucial since it couples the condensate and 
non-condensate degrees of freedom.

The kinetic equation (\ref{eq:kinht}) is valid 
only in the semiclassical limit: it assumes that 
the thermal energy is much greater than the spacing 
between the trap SHO energy levels ($k_{B}T\gg \hbar 
\omega_{0}$, where $\omega_{0}$ is the harmonic 
well frequency) and the average interaction energy 
($k_{B}T\gg gn$). ZNG have given a detailed 
derivation \cite{zngjltp} of (\ref {eq:kinht}) 
at finite temperatures using the approach of Kirkpatrick 
and Dorfman \cite{kd}, who originally considered 
a uniform gas. ZNG have used this kinetic equation 
to derive generalized two-fluid hydrodynamic equations 
of the kind first discussed by Landau in 1941.

Kane and Kadanoff (KK) \cite{kk} first used the 
Kadanoff-Baym (KB) formalism\cite{kb} as a microscopic 
basis for the derivation of the Landau two-fluid 
hydrodynamic equations for a uniform Bose fluid.
In the present work, we use the KB approach to 
derive the kinetic equation ($\ref{eq:kinht}$) for 
a trapped Bose gas, obtaining the same collision 
integrals as derived by ZNG. This is the first step 
in more systematic derivation of kinetic equations 
for a trapped Bose gas. A more detailed and extended 
account of the present calculations
will be published elsewhere\cite{itgnew}.

\section{Equations of motion for non-equilibrium 
Green's functions} 
In terms of quantum field operators, the many-body 
Hamiltonian ($\hat{K}=\hat{H}-\mu_{0}\hat{N}$) 
describing interacting Bosons confined 
by an external harmonic potential 
$U_{ext}({\bf r})$ is given by:
\ba
\hat{K}&=&\int d{\bf r} {\psi}^{\dag}(\r)
\left[-\frac{1}{2m}\nabla^{2}_{\r}+U_{ext}(\r)
-\mu_{0}\right]{\psi}(\r) \nonumber \\
&+&\frac{1}{2} \int d\r d\r'
{\psi}^{\dag}(\r){\psi}^{\dag}(\r') 
v(\r-\r'){\psi}(\r){\psi}(\r').
\ea
We separate out the condensate part of the field 
operator in the usual fashion\cite{fw} 
\begin{equation}
\psi(\r)=\langle\psi(\r)\rangle_{t}+\tilde{\psi}(\r),
\label{eq:seppsi}
\end{equation}
where $\langle \tilde{\psi}(\r)\rangle=0$ and 
$\langle\psi(\r)\rangle_{t}=\Phi(\r,t) $ is the 
Bose macroscopic wavefunction. The non-condensate 
(or excited atom component) field operators  
$\tilde{\psi}(\r)$ and $\tilde{\psi}^{\dag}(\r)$ 
satisfy the usual Bose commutation relations.

In a Bose-condensed 
system, the finite value of $\Phi(\r,t)$ leads 
to finite values of the off-diagonal 
(or anomalous) propagators 
$\langle\tilde{\psi}(1)\tilde{\psi}(1')\rangle$ 
and 
$\langle\tilde{\psi}^{\dag}(1)
\tilde{\psi}^{\dag}(1')\rangle$. These must be dealt
with on an equal basis with the diagonal (or normal) 
propagators, and thus we must work with a $2\times 2$ 
matrix single-particle Green's function 
defined by \cite{kk,hm}
\begin{equation}
\hat{g}(1,1';U)=-i\left( \begin{array}{cc}\langle T 
\psi(1) \psi^{\dag}(1')\rangle\hspace*{5mm}\langle T 
\psi(1) \psi(1')\rangle\\ 
\langle T \psi^{\dag}(1) \psi^{\dag}(1')\rangle
\hspace*{5mm} \langle T\psi^{\dag}(1)\psi(1')\rangle 
\end{array}\right). \label{eq:deftildeg}
\end{equation}
Here, {\it T} represents the time-ordering 
operation and we use the usual KB abbreviated notation, 
1$\equiv (\r,t)$ and $1'\equiv (\r',t'$). We 
define $\hat{g}^{<}$ and $\hat{g}^{>}$ by
\ba
\hat{g}(1,1';U)&=&\hat{g}^{>}(1,1';U)\hspace{10mm}
t_{1}>t_{1'} \nonumber \\
&=&\hat{g}^{<}(1,1';U)\hspace{10mm}   t_{1}<t_{1'}.
\ea
Using (\ref{eq:seppsi}), the matrix propagator in 
(\ref{eq:deftildeg})  splits into two parts
\be
\hat{g}(1,1';U)=\hat{\tilde{g}}(1,1';U)+\hat{h}(1,1';U).
\ee
Here $\hat{\tilde{g}}$ is identical to 
(\ref{eq:deftildeg}) except that it involves the 
non-condensate part of the field operators, while 
the non-condensate part is given by
\be
\hat{h}(1,1';U)\equiv -i\left ( \begin{array}{cc} 
\Phi(1)\Phi^{*}(1')   \hspace*{5mm} \Phi(1) 
\Phi(1') \\ \Phi^{*}(1)\Phi^{*}(1')
\hspace*{5mm} \Phi^{*}(1)\Phi(1') 
\end{array} \right ),
\ee
with $\langle\psi^{\dag}(\r)\rangle_{t}
\equiv \Phi^{*}(\r,t)$. 

A very convenient way of generating the equations
of motion for $\hat{\tilde{g}}$
and $\Phi$ is to use functional derivatives with 
respect to weak external fields \cite{kb,hm}. 
The latter are described by 
\be
H'(t_{1})=\frac{1}{2}\int d{\bf r}_{1} 
d2{\psi}^{\dag}(1)U(1,2){\psi}(2) 
+\int d {\bf r}_{1}\left[\psi^{\dag}(1)\eta_{ext}(1)
+\psi(1)\eta^{*}_{ext}(1)\right],
\label{eq:externalh}
\ee
where $U(1,2) $ is an external generating scalar 
field non-local in space and time. 
It represents a perturbation in which a particle 
is removed from the system at point 1 and 
added at 2. The symmetry-breaking fields 
$\eta_{ext}$ and $\eta^{*}_{ext}$ describe 
particle creation and destruction \cite{hm,bog}. 
Higher-order Green's functions can all be expressed 
as functional derivatives of single-particle 
Green's functions with respect to these fields.

It is useful to define the matrix inverse 
of $\hat{\tilde{g}}$ by
\be
\hat{\tilde{g}}^{-1}(1,1';U)\equiv \hat{g}^{-1}_{0}(1,1')
-U(1,1')-\hat{\Sigma}(1,1';U).
\label{eq:tildeginv}
\ee  
Following the  Kane-Kadanoff (KK) analysis  \cite{kk,kb}, 
the Dyson-Beliaev equations of motion  for the real-time 
non-condensate propagators  $\hat{\tilde{g}}(1,1')$ 
can be conveniently written in the following 
$2 \times 2$ matrix form
\ba 
&&\int d\bar{1}\left[{\hat{g}_{0}}^{-1}(1,\bar{1})-
\hat{\Sigma}^{HF}(1, \bar{1})\right] 
\hat{\tilde{g}}^{\up}(\bar{1},1')\nonumber \\
 &=&\int_{-\infty}^{t_{1}}d\bar{1}\hat{\Gamma}(1,\bar{1})
\hat{\tilde{g}}^{\up}(\bar{1},1')
-\int_{-\infty}^{t_{1'}} d\bar{1}\hat{\Sigma}^{\up}
(1,\bar{1})\hat{a}(\bar{1},1'),
\label{eq:tildeg2}
\ea 
and 
\ba 
&&\int d\bar{1}\hat{\tilde{g}}^{\up}(1,\bar{1})
\left[{\hat{g}_{0}}^{-1}(\bar{1},1')-
\hat{\Sigma}^{HF}( \bar{1},1')\right] \nonumber \\
&=&\int_{-\infty}^{t_{1}}d\bar{1}\hat{a}(1,\bar{1})
\hat{\Sigma}^{\up}(\bar{1},1')
-\int_{-\infty}^{t_{1'}}d\bar{1}\hat{\tilde{g}}^{\up}
(1,\bar{1})\hat{\Gamma}(\bar{1},1').
\label{eq:tildeg3}
\ea 
Here $\hat{a}(1,1')$ and $\hat{\Gamma}(1,1')$ are 
defined by the matrix elements
\ba
&&a_{\al \bt}(1,1')\equiv 
\tilde{g}_{\al \bt}^{>}(1,1')
-\tilde{g}_{\al \bt}^{<}(1,1') \nonumber \\
&&\Gamma_{\al \bt}(1,1')\equiv 
\Sigma_{\al \bt}^{>}(1,1')
-\Sigma_{\al \bt}^{<}(1,1').
\label{eq:defagamma}
\ea
The spectral density $a_{\al \bt}(1,1')$ will play 
a crucial role in our later discussion. In the above 
equations and elsewhere, integration over $d{\bar 1}$ 
means integration  over the coordinates 
$({\bf r}_{1}, t_{1})$ and a trace over the 
matrix index $\alpha_{1}$; and
$\delta(11')\equiv\delta({\r}-{\r'})\delta(t-t')$.

We have explicitly split the self-energy which 
is involved in (\ref{eq:tildeg2}) and (\ref{eq:tildeg3})
into two parts \cite{kk,hm}
\be
\hat{\Sigma}(1,1')=\hat{\Sigma}^{HF}(1,1')
+\hat{\Sigma}_{c}(1,1').
\ee
The Hartree-Fock self-energy is given by, 
using (\ref {eq:continteraction})
\be
\hat{\Sigma}^{HF}(11')=g\left ( \begin{array}{cc} 
2n(1),  \hspace*{3mm} m(1) \\ 
m^{*}(1),\hspace*{3mm} 2n(1) \end{array} 
\right )\delta(11'), \label{eq:sigmahf}
\ee
and $\Sigma_{c}$ is the ``collisional'' part of the 
self-energy. The total density is given by 
$n(1)\equiv i\tilde{g}^{<}_{11}(1,1^{+})+|\Phi(1)|^{2}
=\tilde{n}(1)+n_{c}(1)$ and the total anomalous density by 
$m(1)\equiv i\tilde{g}_{12}(1,1)+[\Phi(1)]^{2}$. 
In addition, we define
\ba
\hat{\Sigma}_{c}(1,1')&=&\hat{\Sigma}^{>}(1,1')
\hspace{10mm} t_{1}>t_{1'} \nonumber \\
&=&\hat{\Sigma}^{<}(1,1')\hspace{10mm}t_{1}<t_{1'}.
\ea 
In (\ref{eq:tildeg2}) and (\ref{eq:tildeg3}), 
the inverse of the $2\times 2$ matrix non-interacting
Bose gas propagator $\hat{g}_{0}(1,1')$ is defined by
\be
\hat{g}_{0}^{-1}(1,1')=\left[i{\bf \tau}_{3}
\frac{\partial}{\partial t_{1}}
+\frac{\nabla_{1}^{2}}{2m}-U_{ext}(\r_{1})+\mu_{0}
\right] \delta(1,1').
\label{eq:godef}
\ee
 
We note that the equations in (\ref{eq:tildeg2}) 
and (\ref{eq:tildeg3}) already have the ``structure'' 
of the kinetic equation in (\ref{eq:kinht}). 
The Hartree-Fock part of the self-energy is 
included into the left-hand side of (\ref{eq:tildeg2}) 
and (\ref{eq:tildeg3}), giving the mean-field 
contribution to the ``streaming'' term. The second 
order self-energy describing binary collisions is 
included on the right-hand side of (\ref{eq:tildeg2}) and 
(\ref{eq:tildeg3}) and it will eventually be shown to 
give rise to the collision integrals in (\ref{eq:collint}).  

The specific form of $\Sigma_{c}$ will depend on 
the approximation that we use. In this paper, 
we work with the second-order self-energy given by 
the Beliaev (gapless) approximation \cite{hm,hfb}. 
The advantage of the Beliaev approximation is that 
the non-condensate Green's function exhibits the 
correct spectrum (phonon-like in the 
long-wavelength, uniform gas limit). 
In contrast, ``conserving approximations'' 
are based on a functional, from which both self-energy 
$\Sigma$ and the source $\eta$ functions  can be 
derived by functional differentiation. The resulting 
single-particle Green's function can be used to 
generate a density response function whose spectrum 
is guaranteed to satisfy conservation 
laws \cite{hm,hfb,kbphyrev,baym}, even though 
the Green's function has an energy gap in the 
long-wavelength limit. Kinetic equations 
for a trapped Bose gas based on a 
``conserving approximation'' will be derived 
elsewhere \cite{itgnew}.

The equation for the condensate can be 
written as \cite{hm}
\be
\int d\bar{1}{\hat{g}_{0}}^{-1}(1,\bar{1})
\hat{G}_{1/2}(\bar{1})=\sqrt{-i}\hat{\eta}(1)
+\sqrt{-i}\hat{\eta}_{ext}(1),
\label{eq:orderg1/2}
\ee 
where the so-called condensate source function 
$\eta$ is defined by the three-field 
correlation function
\be
\sqrt{-i}\hat{\eta}(1) \equiv \frac{1}{2}
\int d2 \sqrt{-i}v(12)\langle
T\Psi(1)\Psi^{\dag}(2)\Psi(2)\rangle.
\label{eq:etadef}
\ee
Here $\Psi$ is defined as
\be
\Psi(1)\equiv \left(\begin{array}{c} \psi(1) \\  
\psi^{*}(1) \end{array}\right). 
\ee
We have introduced the 2-component order parameter 
$\hat{G}_{1/2}(1)\equiv\sqrt{-i}\langle\Psi(1)\rangle$, 
and  $\hat{\eta}^{ext}$ describes the external 
particle-source fields defined in (\ref{eq:externalh}), 
with
\be
\hat{\eta}_{ext}(1) \equiv \left( \begin{array}{c} 
\eta_{ext}(1) \\ 
\eta^{*}_{ext}(1) \end{array}\right ).
\ee
The external generating fields $U$ and $\eta_{ext}$ 
will be left implicit in the rest of this paper.

\section{Beliaev approximation for the self-energy}

The self-energy in the Beliaev (gapless) approximation 
is {\it defined} by \cite{hm}
\be
\sqrt{-i}\frac{\delta \hat{\eta}(1)}
{\delta \hat{\tilde{G}}_{1/2}(1')}
\equiv \hat{\Sigma}(11').
\label{eq:sigmabel}
\ee
To derive the second-order 
Beliaev approximation for the self-energy, we need 
to find the second-order expression for 
the source function $\hat{\eta}$. For clarity, in 
this section we will use a general interatomic potential 
{\it v}, rather than the {\it s}-wave approximation 
given by (\ref {eq:continteraction}).
We express the three-point function in 
(\ref{eq:etadef}) in terms of functional derivatives 
in the following way \cite{hm,martin}
\be
\sqrt{-i}\hat{\eta}(1)=iv(12) 
\left[\frac{1}{2}\left[\hat{\tilde{G}}(22)
+\hat{G}_{1/2}(2)\hat{G}_{1/2}^{\dag}(2)
\right]\hat{G}_{1/2}(1)+i\frac{\delta \hat{G}_{1/2}(1)}
{\delta U(22)}\right],
\label{eq:etagen}
\ee
where we recall that repeated matrix labels are 
summed over. $\hat{\tilde{G}}(1,1')$ is the 
imaginary-time non-condensate propagator, 
from which one obtains the real-time propagator 
$\hat{\tilde{g}}(1,1')$ used in the KB formalism. 
Equation (\ref{eq:etagen}) can also be written as 
(from now on, the matrix nature of the quantities 
is left implicit):
\ba
\sqrt{-i}\eta(1)&=&iv(12) 
\left[\frac{1}{2}\left[\tilde{G}(22)
+G_{1/2}(2)G_{1/2}^{\dag}(2)\right]
G_{1/2}(1)+\tilde{G}(12)G_{1/2}(2)\right]
\nonumber \\
&+&\frac{i}{2}v(12)\tilde{G}(14)\tilde{G}(23)
\frac{\delta \Sigma(35)}{\delta 
G_{1/2}^{\dag}(4)}\tilde{G}(52).
\label{eq:etasecond}
\ea
We approximate $\Sigma$ in (\ref{eq:etasecond}) by the 
full Hartree-Fock-Bogoliubov (HFB) first-order 
self-energy (denoted as HF for simplicity) 
\ba
\Sigma_{HF}(11')&=&\frac{i}{2}v(12)
\left[ G_{1/2}(2)G_{1/2}^{\dag}(2)
+\tilde{G}(22)\right]\delta(11') \nonumber \\
&+&iv(11') \left[ G_{1/2}(1)G_{1/2}^{\dag}(1')
+\tilde{G}(11')\right]. \label{eq:sigmaga}
\ea
If one recalls that $G_{1/2}(1)\equiv \sqrt{-i}\Phi(1)$ 
and the definitions for {\it m}(1) and {\it n}(1) given 
below (\ref{eq:sigmahf}), one can immediately see that 
for the contact interaction (\ref {eq:continteraction}), 
(\ref{eq:sigmaga}) reduces to (\ref{eq:sigmahf}). 
Using (\ref{eq:sigmaga}), we obtain the following 
second-order expression for the source function 
\ba
\sqrt{-i}\eta(1)&=&\frac{i}{2}v(12)
\left[G_{1/2}(1)G_{1/2}(2)G_{1/2}^{\dag}(2)
+G_{1/2}(1)\tilde{G}(22)\right] \nonumber \\
&+&iv(12)G_{1/2}(2)\tilde{G}(12) 
-\frac{1}{2}v(13)v(24)\tilde{G}(32)
\tilde{G}(23)\tilde{G}(14)G_{1/2}(4) \nonumber \\
&-&v(13)v(24)\tilde{G}(14)
\tilde{G}(43)\tilde{G}(32)G_{1/2}(2).
\label{eq:condbeliaev}
\ea
To obtain a second-order expression for the self-energy, 
we use (\ref{eq:condbeliaev}) in (\ref{eq:sigmabel}). 
In this gapless approximation, we note that 
$\tilde{G}$ and $G_{1/2}$ are not independent; using  
(\ref{eq:tildeginv}), one finds
\be
\frac{\delta \tilde{G}}{\delta G_{1/2}}=\tilde{G}
\frac{\delta \Sigma}{\delta G_{1/2}}\tilde{G}.
\ee
Using (\ref{eq:sigmabel}), we obtain the second 
order Beliaev collisional part of the self-energy 
\ba
\hat{\Sigma}_{c}(1,1')=&-&\frac{1}{2}v(13)v(21')
\tilde{G}(11')\left[\tilde{G}(23)\tilde{G}(32)
+\tilde{G}(23)h(32)+h(23)\tilde{G}(32)\right]
\nonumber \\
&-&v(13)v(21')\tilde{G}(12)\left[\tilde{G}(23)h(31')+
h(23)\tilde{G}(31')+\tilde{G}(23)\tilde{G}(31')
\right] \nonumber \\
&-&\frac{1}{2}v(13)v(21')\left[h(11')\tilde{G}(23)
\tilde{G}(32)+2h(12)\tilde{G}(23)\tilde{G}(31')\right].
\label{eq:sigmabeliaev}
\ea
We now use these results to find, in a consistent manner,
explicit expressions for the collision integrals in our 
kinetic equation and an equation of motion for the 
condensate \cite{itgnew}.

\section{Generalized kinetic equations}

To derive a generalized kinetic equation, we first 
gauge transform (\ref{eq:tildeg2}) and (\ref{eq:tildeg3}) 
to the local rest frame to remove the phase of the 
macroscopic wavefunction \cite{kk}. They remain 
unchanged in form when $g_{0}^{-1}$ is replaced by 
(compare with (\ref{eq:godef}))
\be
{g}_{0}^{-1}(1,1')=\left[ i {\bf \tau}_{3}
\frac{\partial}{\partial t_{1}}-\frac{\partial \theta(1)}
{\partial t_{1}}+\frac{1}{2}[\nabla_{1}+i{\bf \tau}_{3}
\nabla_{1}\theta(1) ]^{2}-U_{ext}(\r_{1})+\mu_{0} \right]
 \delta(1,1'). \label{eq:g0llf}
\ee
We recall that the superfluid velocity ${\bf v}_{s}\co$ 
and the local chemical potential $\mu\co$ are 
defined by \cite{kk}
\ba
m{\bf v}_{s}\co&\equiv& \nabla_{\bf R}\theta\co     
\nonumber \\
\frac{\partial \theta\co}{\partial T}
&\equiv&-\left[\mu\co-\mu_{0}
+\frac{1}{2}m {v_{s}}^{2}\co\right]. 
\label{eq:defvsmu}
\ea
where, in the lab frame, the condensate wavefunction 
is given by $\Phi\co\equiv \sqrt{n_{c}\co}e^{i\theta\co}$.
 
To illustrate how this method works, in this brief report 
we neglect the off-diagonal anomalous Green's functions 
($\tilde{g}_{12}$ and $\tilde{g}_{21}$). 
The same approach can be used when we include the 
off-diagonal propagators \cite{itgnew}. 
In this approximation, the $2\times 2$ matrix equations 
in (\ref{eq:tildeg2}) and (\ref{eq:tildeg3}) reduce 
to scalar equations for the $\tilde{g}_{11}^{\up}$ 
component
\ba 
&&\left[i\frac{\partial}{\partial 
t_{1}}-\frac{\partial \theta(1)}{\partial t_{1}}
+\frac{1}{2}[\nabla_{1}+im{\bf v}_{s}(1)]^{2}
-U(1)+\mu_{0}\right] \tilde{g}_{11}^{\up}
(\bar{1},1')  \nonumber \\
&&=\int_{-\infty}^{t_{1}}d\bar{1}\Gamma_{11}(1,\bar{1})
\tilde{g}_{11}^{\up}(\bar{1},1')
-\int_{-\infty}^{t_{1'}} d\bar{1}\Sigma_{11}^{\up}
(1,\bar{1})a_{11}(\bar{1},1')
\label{eq:popov11}
\ea
and
\ba 
&&\left[-i\frac{\partial}{\partial 
t_{1'}}-\frac{\partial \theta(1')}{\partial t_{1'}}
+\frac{1}{2}[\nabla_{1'}-im{\bf v}_{s}(1')]^{2}-U(1')
+\mu_{0}\right] \tilde{g}_{11}^{\up}
(\bar{1},1')\nonumber \\
&&=\int_{-\infty}^{t_{1}}d\bar{1}a_{11}(1,\bar{1})
\Sigma_{11}^{\up}(\bar{1},1')
-\int_{-\infty}^{t_{1'}}d\bar{1}{\tilde{g}}_{11}^{\up}
(1,\bar{1})\Gamma_{11}(\bar{1},1').
\label{eq:popov21}
\ea  
The effective self-consistent Hartree-Fock dynamic mean 
field $U(1)$ is given by (\ref {eq:effectivefield}). 

Following KB \cite{kb}, we find it very useful to 
express correlation functions in terms of relative and 
center-of-mass coordinates defined by
\be
\r=\r_{1}-\r_{1'}, \hspace{5mm} t=t_{1}-t_{1'}; 
\hspace{5mm} {\bf R}=\frac{\r_{1}+\r_{1'}}{2},\hspace{5mm} 
T=\frac{t_{1}+t_{1'}}{2}.
\ee
 Correlation functions (like $\tilde{g}, \Sigma$, etc.) 
 are dominated by the small values of relative 
coordinates $(\r, t)$ (or high momenta and frequencies 
in the Fourier transforms), but vary slowly as  
functions of the center-of-mass coordinates $\co$. 
Using these key properties of correlation functions 
to simplify the equations, we write  
(\ref{eq:popov11}) and (\ref{eq:popov21}) in terms of 
the  center-of-mass and relative coordinates 
to obtain (see Ref. 7 for details)
\ba 
&&\left[i\frac{\partial}{\partial T}+\left[\r
\cdot \nabla_{\bf R}+t\frac{\partial}{\partial T}\right]
\left(\mu\co- U\co\right)+i\nabla_{\bf R} 
\cdot {\bf v}_{s}\co\right. \nonumber \\
&+&\left.\frac{1}{m}\nabla_{\bf R}\cdot 
\nabla_{\r}+i\left[\left(\r \cdot \nabla_{\bf R}+
t\frac{\partial}{\partial T}\right)
{\bf v}_{s}\co\right]\cdot \nabla_{\r}
+i{\bf v}_{s}\co\cdot \nabla_{\bf R}\right]
\tilde{g}_{11}^{\down}(\r,t) \nonumber \\ 
&=&\int d\bar{\r}d\bar{t} 
\left[\tilde{g}_{11}^{>}(\bar{\r},\bar{t})
\Sigma_{11}^{<} (\r-\bar{\r},t-\bar{t};{\bf R},T)
-\tilde{g}_{11}^{<}(\bar{\r},\bar{t})
\Sigma_{11}^{>} (\r-\bar{\r},t-\bar{t};{\bf R},T) 
\right].
\label{eq:popovcm} 
\ea 
The ({\bf R},{\it T} ) dependence of 
$\tilde{g}_{11}(\r,t;{\bf R},T)$ is left implicit.
The double Fourier transform of (\ref{eq:popovcm}) 
is given by 
\ba 
&&\left[\frac{\partial}{\partial T}+
\nabla_\vp\left[\tilde{\epsilon}_{p}
+\vp\cdot {\bf v}_{s}\right]\cdot 
\nabla_{\bf R}-\nabla_{\bf R} 
\left[\nee+\vp\cdot {\bf v}_{s}\right] \cdot 
\nabla_\vp \right. \nonumber \\
&+&\left. \frac{\partial}{\partial 
T}\left[\nee+\vp\cdot {\bf v}_{s}\right] 
\frac{\partial}{\partial\omega}\right] 
\tilde{g}_{11}^{\down}\pomrt \nonumber \\
&=&\tilde{g}_{11}^{>}\pomrt 
\Sigma_{11}^{<}\pomrt-\tilde{g}_{11}^{<}\pomrt
\Sigma_{11}^{>}\pomrt. 
\label{eq:kineqquasi} 
\ea 
Here, we have defined the ``normal'' HF 
single-particle energy by
\be 
\nee\co\equiv \frac{p^{2}}{2m}+U\co-\mu\co.
\label{eq:hfspectrum}
\ee 
In the so-called Thomas-Fermi approximation 
\cite{strtheory}, we note that this energy reduces to
\be 
\nee\co\equiv \frac{p^{2}}{2m}
+gn_{c}\co.
\label{eq:hftf}
\ee  

Using (\ref{eq:kineqquasi}) to calculate the 
spectral density as defined in (\ref{eq:defagamma}), 
we find that the self-energies 
$\Sigma^{\down}$ cancel out and we are left with

\ba
\left[\frac{\partial}{\partial T}\right.&
+&\left.\nabla_{\bf p}\left[\nee
+\vp\cdot {\bf v}_{s}\right]\cdot 
\nabla_{\bf R} - \nabla_{\bf R}\left[\nee
+\vp\cdot {\bf v}_{s}\right] 
\cdot \nabla_\vp\right. \nonumber \\
&+&\left.\frac{\partial}{\partial T}
\left[\nee+\vp \cdot {\bf v}_{s} \right] 
\frac{\partial}{\partial\omega}\right] a_{11}\pomrt=0. 
\label{eq:defa} 
\ea 
This equation will prove very important in 
deriving a kinetic equation for the distribution 
function for the non-condensate atoms. 
One may explicitly verify  that the quasiparticle 
approximation for the spectral density given by
\be 
a_{11}\pomrt=(2\pi)\delta(\omega-\vp\cdot
{\bf v}_{s}-\nee\co)  \label{eq:quasia}
\ee  
satisfies (\ref{eq:defa}). We note that the 
assumption that the correlation functions 
$\tilde{g}^{\down}(\r,t;{\bf R},T) $ are 
peaked at small values of ${\it \r}$ and {\it t} 
implies that large values of {\bf p} and $\omega$ 
determine $\tilde{g}^{\down}\pomrt$, where the 
particle-like spectrum (\ref {eq:hfspectrum}) 
is a good approximation.  

Following Ref \cite{kb}, we introduce a 
quasiparticle distribution function $f$ 
in the following way (see (\ref{eq:defagamma}))
\ba 
&&\tilde{g}_{11}^{<}\pomrt \equiv 
a_{11}\pomrt f({\bf p},{\bf R},T) \nonumber \\ 
&&\tilde{g}_{11}^{>}\pomrt 
\equiv a_{11}\pomrt [1+f({\bf p},{\bf R},T)]. 
\label{eq:deff}
\ea 
We recall that the usual Wigner distribution 
function is related to the KB diagonal 
single-particle Green's function by
\be
f\cof\equiv \int d{\bf r}e^{-i{\bf p}{\bf r}}
i\tilde{g}_{11}^{<}({\bf r}, t=0;{\bf R},T).
\label{eq:wignerf}
\ee 
Using (\ref{eq:deff}) in conjunction with the 
explicit result for $a_{11}$ in (\ref{eq:quasia}), 
we see that the quasiparticle distribution function 
in (\ref{eq:deff}) reduces to the Wigner distribution 
function (\ref {eq:wignerf}). This is correct in the 
high-temperature limit only, where the  HF quasiparticle 
spectrum reduces to the Hartree-Fock single-particle 
excitation spectrum and the quasiparticle distribution 
function is equivalent to 
the particle distribution function.

Using (\ref{eq:defa}), (\ref {eq:quasia}), 
and (\ref{eq:deff}) and after integration over $\omega$, 
(\ref{eq:kineqquasi}) reduces to a kinetic equation 
for the distribution function {\it f}, namely 
\ba 
&&\left[\frac{\partial}{\partial T} 
+\nabla_\vp\left[\nee+\vp\cdot {\bf v}_{s}\right]
\cdot \nabla_{\bf R} - \nabla_{\bf R}\left[\nee+
\vp\cdot {\bf v}_{s}\right] \cdot \nabla_\vp \right]
f\cof \nonumber \\ 
&&=(1+f)\Sigma_{11}^{<}(\vp,\omega=\nee+\vp
\cdot {\bf v}_{s};{\bf R},T)-f\Sigma_{11}^{>}
(\vp,\omega=\nee+\vp\cdot {\bf v}_{s};{\bf R},T), 
\label{eq:quasikin}
\ea
where $f\equiv f\cof$. A generalized quantum 
kinetic equation of this kind has been derived 
by Stoof \cite{stoofjltp} using the related Keldysh 
formalism and path integrals.

\section{Collision integrals in the second-order 
Beliaev approximation}

Neglecting the off-diagonal Green's functions, 
the Fourier transform of the collisional part of 
the self-energy  $\Sigma^{\down}$, given 
by (\ref{eq:sigmabeliaev}), is given by the following 
expression
\ba 
&&\Sigma_{11}^{\down}({\bf p},\omega)=2g^{2}\int 
\frac{d\vp_{i}d\omega_{i}} 
{(2\pi)^{12}}(2\pi)^{4}\delta(\omega+\omega_{1}
-\omega_{2}-\omega_{3}) 
\delta(\vp+\vp_{1}-\vp_{2}-\vp_{3})\nonumber \\ 
&&\times \left[\tilde{g}_{11}^{\up}(\vp_{1},\omega_{1}) 
\tilde{g}_{11}^{\down}(\vp_{2},\omega_{2}) 
h_{11}(\vp_{3},\omega_{3}) + 
\tilde{g}_{11}^{\up}(\vp_{1},\omega_{1}) 
h_{11}(\vp_{2},\omega_{2}) 
\tilde{g}_{11}^{\down}(\vp_{3},\omega_{3}) \right.  
\nonumber \\ 
&+&\left.  h_{11}(\vp_{1},\omega_{1}) 
\tilde{g}_{11}^{\down}(\vp_{2},\omega_{2}) 
\tilde{g}_{11}^{\down}(\vp_{3},\omega_{3}) 
+\tilde{g}_{11}^{\up}(\vp_{1},\omega_{1}) 
\tilde{g}_{11}^{\down}(\vp_{2},\omega_{2}) 
\tilde{g}_{11}^{\down}(\vp_{3},\omega_{3})\right] 
\label{eq:fsigmacons}
\ea 
Here and elsewhere, the ({\bf R},{\it T} ) 
dependence of the self-energies 
and the Green's functions is left implicit.
Using (\ref{eq:quasia}), (\ref{eq:quasikin}), and 
(\ref{eq:fsigmacons}), one can evaluate the collision 
integral $C_{22}$ which involves collisions 
between excited atoms (i.e., the parts of 
the self-energy in (\ref{eq:fsigmacons}) which 
involve three propagators for 
non-condensate atoms). We obtain \cite{itgnew}
\ba
C_{22}[f]&=&2g^{2}\int \frac{d\vp_{1} d\vp_{2} 
d{\bf p}_{3}}{(2\pi)^{3}}\delta(\nee+
\tilde{\epsilon}_{p_{1}}-\tilde{\epsilon}_{p_{2}}
-\tilde{\epsilon}_{p_{3}})
\delta({\bf p}+{\bf p}_{1}-{\bf p}_{2}-{\bf p}_{3})
\nonumber \\
&&\times \left[ (1+f)(1+f_{1})f_{2}f_{3}-f 
f_{1}(1+f_{2})(1+f_{3})\right],
\label{eq:c22}
\ea
where $f\equiv f\cof$ and $f_{i}\equiv 
f({\bf p}_{i},{\bf R},T)$. $C_{22}$ describes 
collisions where 2 excited atoms are scattered 
into 2 excited atoms, hence the subscript 22.

The collision integral $C_{12}[f]$ involving collisions 
between normal atoms and one condensate atom 
comes from the terms involving $h_{11}$ in 
(\ref{eq:fsigmacons}). One can show that for 
slowly varying external disturbances \cite{kk}, 
one has
\be
h_{11}\pomrt =n_{c}\co\delta(\vp)
\delta(\omega)(2\pi)^{4}. \label{eq:hfourier}
\ee
Using (\ref{eq:hfourier}), $C_{12}[f]$ is found 
to be given by \cite{itgnew}    
\ba 
&&C_{12}[f]=4\pi g^{2}n_{c}\co
\int \frac{d\vp_{1} d\vp_{2} d\vp_{3}}
{(2\pi)^{3}}\left[ \delta({\bf p}-{\bf p}_{1})
-\delta({\bf p}-{\bf p}_{2})
-\delta({\bf p}-{\bf p}_{3})\right]
\nonumber \\
&&\delta(\tilde{\epsilon}_{p_{1}}
-\tilde{\epsilon}_{p_{2}}-\tilde{\epsilon}_{p_{3}})
\delta({\bf p}_{1}-{\bf p}_{2}-{\bf p}_{3})
\left[(1+f_{1})f_{3}f_{2}-f_{1}(1+f_{3})
(1+f_{2})\right]. \label{eq:c12}
\ea
$C_{12}$ describes collisions where we go from 
1 excited and one condensate atom to 2 excited 
atoms, hence the subscript 12.
Inserting (\ref{eq:c22}) and (\ref{eq:c12}) 
into (\ref{eq:quasikin}), we arrive at a kinetic 
equation identical to that derived by 
ZNG \cite{zngjltp}, namely
\ba 
&&\left[\frac{\partial}{\partial T}+\nabla_\vp
\left[\nee+\vp\cdot {\bf v}_{s}\right]\cdot 
\nabla_{\bf R} -\nabla_{\bf R}\left[\nee+\vp
\cdot {\bf v}_{s}\right] \cdot \nabla_\vp 
\right] f\cof \nonumber \\ 
&=&C_{22}[f\cof]+C_{12}[f\cof].
\label{eq:eqkinht}
\ea
The left-hand side of (\ref{eq:eqkinht}) is 
the expected streaming term in the frame of 
reference in which the superfluid velocity is 
zero (ZGN have derived the equivalent of 
(\ref{eq:eqkinht}) in the 
lab frame). For a uniform gas at finite temperatures, 
(\ref {eq:eqkinht}) was first obtained 
by Kirkpatrick and Dorfman \cite{kd} using a 
completely different approach. 

\section{Generalized Gross-Pitaevskii equation}
To derive an equation of motion for the order 
parameter, we first write equation (\ref {eq:orderg1/2}) 
for $\Phi(\r,t)$ in the local rest frame, namely
\ba
\left[i\frac{\partial}{\partial t}
\right.&-&\left.\frac{\partial \theta(1)}
{\partial t}+\frac{1}{2m}\left[\nabla_{\r}
+im{\bf v}_{s}(1)\right]^{2}+
\mu_{0}-U_{ext}(\r)-g\left(2\tilde{n}(1)+n_{c}(1)
\right)\right]\Phi(1) \nonumber \\
&=&\int_{-\infty}^{t}d\bar{1}\left[S^{>}_{11}
-S^{<}_{11}\right]\left(\r-\bar{\r},t-\bar{t};
(\r+\bar{\r})/2,(t+\bar{t})/2\right)
\Phi(\bar{\r},\bar{t}). \label{eq:orderpa}
\ea
We assume, as usual, that the $S_{11}$ correlation 
function related to (\ref{eq:etadef}) is 
dominated by small values of the relative coordinates
$(\r-\bar{\r},t-\bar{t})$, and therefore we can 
approximate $S^{\down}_{11}$ in (\ref{eq:orderpa}) by 
$S^{\down}(\r-\bar{\r},t-\bar{t};\r,t)$. For 
the same reason, we can approximate the macroscopic 
wavefunction in (\ref{eq:orderpa}) by 
$\Phi(\bar{1})\simeq \Phi(1)\equiv \sqrt{n_{c}(\r,t)}$. 
Hence, we approximate (\ref{eq:orderpa}) by
\ba
\left[i\frac{\partial}{\partial t}
\right.&-&\left. \frac{\partial \theta(1)}
{\partial t}+\frac{1}{2m}\left[ \nabla_{\r}
+im{\bf v}_{s}(1)\right]^{2}+\mu_{0}-
U_{ext}(\r)-g\left(2\tilde{n}(1)+n_{c}(1)
\right)\right]\Phi(1) \nonumber \\
&=&\int_{-\infty}^{t}d\bar{\r}d\bar{t}
\left[S^{>}_{11}-S^{<}_{11}\right]
\left(\r-\bar{\r},t-\bar{t};\r,t\right)
\Phi(\r,t). \label{eq:orderexact}
\ea
We can rewrite (\ref{eq:orderexact}) as 
(renaming $(\r,t)\rightarrow \co$)
\ba
&&\left[i\frac{\partial}{\partial T}-
\frac{\partial \theta\co}{\partial T}
+\frac{1}{2m}\left[ \nabla_{\bf R}
+im{\bf v}_{s}\co\right]^{2}
+\mu_{0}-U_{ext}({\bf R}) \right. \nonumber \\
&-&\left.g\left(2\tilde{n}\co+n_{c}\co\right)
\right]\Phi\co\nonumber \\
&=&\Phi\co\int\frac{d{\bf p}d\omega}{(2\pi)^{4}}
\left[S^{>}_{11}-S^{<}_{11}\right]\pomrt
\int_{-\infty}^{T}d\bar{\r} d\bar{t} 
e^{i{\bf p}({\bf R}-\bar{\r})-i\omega (T-\bar{t})}
 \label{eq:orderexact1}
\ea
In the Beliaev approximation, $S_{11}\pomrt$ 
is found to be given by \cite{itgnew}
\ba
S_{11}^{\down}({\bf p},\omega;{\bf R},T)&
=&2g^{2}\int \frac{d{\bf p}_{i}d\omega_{i}}
{(2\pi)^{12}}(2\pi)^{4}\delta(\omega+\omega_{1}
-\omega_{2}-\omega_{3})\delta({\bf p}
+{\bf p}_{1}-{\bf p}_{2}-{\bf p}_{3}) 
\nonumber \\
&\times &\tilde{g}_{11}^{\up}
({\bf p}_{1},\omega_{1};{\bf R},T)
\tilde{g}_{11}^{\down}
({\bf p}_{2},\omega_{2};{\bf R},T)
\tilde{g}_{11}^{\down}
({\bf p}_{3},\omega_{3};{\bf R},T). 
\label{eq:conservings}
\ea
In evaluating the right-hand side of 
(\ref{eq:orderexact1}), we have used the following identity
\be
\lim_{\delta \rightarrow 0^{+}}
\int_{-\infty}^{T}d\bar{t} 
e^{-i(\omega+i\delta )(T-\bar{t})}\simeq \pi 
\delta(\omega)+iP\left(\frac{1}{\omega}\right),
\ee 
in conjunction with (\ref{eq:quasia}) and 
(\ref{eq:deff}). We finally obtain a generalized 
Gross-Pitaevskii equation in the following form
\ba
&&i\frac{\partial \sqrt{n_{c}\co}}{\partial T}=
\left[\frac{\partial \theta\co}{\partial T}
-\frac{1}{2m}\left[ \nabla_{\bf R}+im{\bf 
v}_{s}\co\right]^{2}-\mu_{0}\right. \nonumber \\
&+&\left.U_{ext}({\bf R})+g\left[
2\tilde{n}\co+n_{c}\co\right]-iR\co
\right]\sqrt{n_{c}\co}. 
\label{eq:gengp}
\ea
The new dissipative term {\it R} in the GP 
equation is related to the $C_{12}$ collision 
term in the kinetic equation (\ref {eq:eqkinht}), 
namely \cite{itgnew,zngjltp}
\be
R\co\equiv \int \frac{d{\bf p}}{(2\pi)^{3}}
\frac{C_{12}[f\cof]}{2n_{c}\co}.
\ee
This term describes the damping of condensate 
amplitude fluctuations due to collisions with 
the atoms in the thermal cloud. The appearance 
of the dissipative term in (\ref{eq:gengp}) is 
expected since the $C_{12}$ collisions change 
the number of atoms in the condensate and hence 
modify the magnitude of the condensate macroscopic 
wavefunction. The real part of the right-hand side of 
(\ref{eq:orderexact1}) is omitted since we only work 
to first order in the interaction as far as 
renormalized energies are concerned.  
We note that if we transform back into the lab 
frame (recall that in the lab frame 
$\Phi=\sqrt{n_{c}}e^{i\theta}$), (\ref{eq:gengp}) 
reduces to the time-dependent generalized 
Gross-Pitaevskii equation for $\Phi\co$
discussed by ZNG.

\section{Concluding remarks}
In this brief report, we have shown how the elegant 
and powerful Kadanoff-Baym formalism can be used 
in a trapped Bose gas to derive a generalized 
Gross-Pitaevskii equation for the 
condensate wavefunction as well as a quantum 
kinetic equation for distribution function for 
the non-condensate atoms\cite{itgnew}. 

We have limited ourselves to discussing the case 
of finite temperatures, where the single-particle 
spectral density can be approximated by 
(\ref{eq:quasia}). In another paper, we will show 
how this formalism can be extended to deal with the 
case of low temperatures, where both 
the diagonal and the off-diagonal components of the
spectral density $a_{\al \bt}$ must be kept and 
they exhibit the Bogoliubov-Popov spectrum in place of the 
particle-like spectrum in (\ref{eq:quasia}).
 
\acknowledgements
AG would like to acknowledge his intellectual debt 
to Leo Kadanoff and Gordon Baym. Their classic book 
has been a constant inspiration for 
my own work over many decades. 
This research was supported by NSERC.

\end{document}